\documentclass[11pt,reqno]{amsart}
\usepackage{amsmath}
\usepackage{amssymb}
\usepackage{amsthm,amsxtra}
\newtheorem{cor}{Corollary}
\newtheorem{lemma}{Lemma}
\newtheorem{prop}{Proposition}
\newcommand{\BB}{\ensuremath{V_{(0,1)}^{ \mathrm{\mathrm{hard}}}}}

\begin{document}
\title[Relationships between $\tau$-function
and Fredholm determinant ]{Relationships between $\tau$-function and
Fredholm determinant expressions for gap probabilities in random
matrix theory}

\author{Patrick  Desrosiers}
\address{Department of Mathematics and Statistics, University of
Melbourne, Parkville, Victoria 3010, Australia.}
\email{P.Desrosiers@ms.unimelb.edu.au}
\author{Peter J. Forrester}
\address{Department of Mathematics and Statistics, University of
Melbourne, Parkville, Victoria 3010, Australia.}
\email{P.Forrester@ms.unimelb.edu.au}

\date{April 2006}

\begin{abstract}

The gap probability at the hard and soft edges of scaled random
matrix ensembles with orthogonal symmetry are known in terms of
$\tau$-functions. Extending recent work relating to the soft edge,
it is shown that these $\tau$-functions, and their generalizations
to contain a generating function parameter, can be expressed as
Fredholm determinants. These same Fredholm determinants occur in
exact expressions for the same gap probabilities in scaled random
matrix ensembles with unitary and symplectic symmetry.
\end{abstract}

\maketitle
\section{Introduction}

In the 1950's Wigner introduced random real symmetric matrices to
model the highly excited energy levels of heavy nuclei (see
\cite{Po65}). From the experimental data, a natural statistic to
calculate empirically is the distribution of the spacing between
consecutive levels, normalized so that the spacing is unity. For
random real symmetric matrices $X$ with independent Gaussian entries
such that the joint probability density function (p.d.f.) for the
elements is proportional to $e^{-{\rm Tr}(X^2)/2}$ (such matrices
are said to form the Gaussian orthogonal ensemble, abbreviated GOE),
Wigner used heuristic reasoning to surmise that the spacing
distribution is well approximated by the functional form
\begin{equation}\label{pW}
p_1^{\rm W}(s) := {\pi s \over 2} e^{- \pi s^2/4}.
\end{equation}
In the limit of infinite matrix size, it was subsequently proved by
Gaudin that the exact spacing distribution is given by
\begin{equation}\label{pd}
p_1(s) = {d^2 \over d s^2} \det (\mathbb{I} - K_{(0,s)}^{\rm bulk,
+} ),
\end{equation}
where $\mathbb{I}$ stands for the identity operator and where
$K_{(0,s)}^{\rm bulk, +}$ is the integral operator supported on
$(0,s)$ with kernel
\begin{equation}\label{pd1}
{\sin \pi (x - y) \over  \pi (x - y) }
\end{equation}
restricted to its even eigenfunctions. It was shown that this
integral operator commutes with the differential operator for the so
called prolate spherical functions, and from the numerical
determinantion of the corresponding eigenvalues (\ref{pd}) was
computed and shown to differ from the approximation (\ref{pW}) by no
more than a few percent.

The (Fredholm) determinant in (\ref{pd}) is itself a probabilistic
quantity. Thus let $E_1^{\rm bulk}(0;(0,s))$ denote the probability
that for the infinite  GOE, scaled so that the mean spacing is
unity, the interval $(0,s)$ of the spectrum contains no eigenvalues.
Then
\begin{equation}\label{S1}
E_1^{\rm bulk}(0;(0,s)) = \det ( \mathbb{I} - K_{(0,s)}^{\rm bulk,
+} ).
\end{equation}

In applications of random matrices to the eigenspectrum of quantum
Hamiltonians, two other ensembles in addition to the GOE are
relevant. These are the Gaussian unitary ensemble (GUE) of complex
Hermitian matrices, and the Gaussian symplectic ensemble (GSE) of
Hermitian matrices with real quaternion elements. For the infinite
limit of such ensembles of matrices, scaled so that the mean density
is unity, let $E_2^{\rm bulk}(0;(0,s))$ and $E_4^{\rm
bulk}(0;(0,s))$ respectively denote the probabilities that the
interval $(0,s)$ is free of eigenvalues. Then it is known
\cite{Dy62,DM} that
\begin{align}\label{1.5}
E_2^{\rm bulk}(0;(0,s)) & =  \det ( \mathbb{I} - K_{(0,s)}^{\rm bulk} ) \nonumber \\
& =   \det ( \mathbb{I} - K_{(0,s)}^{\rm bulk,+} )
 \det ( \mathbb{I} - K_{(0,s)}^{\rm bulk,-} )
\end{align}
while
\begin{equation}\label{S3}
E_4(0;(0,s)) = {1 \over 2} \Big ( \det ( \mathbb{I} -
K_{(0,2s)}^{\rm bulk, +} ) +
 \det ( \mathbb{I} - K_{(0,2s)}^{\rm bulk, -} ) \Big )
\end{equation}
where $K^{\rm bulk, -}_J$ denotes the integral operator $K^{\rm
bulk}_J$ restricted to odd eigenfunctions.

The remarkable structure exhibited by (\ref{S1})--(\ref{S3}) can
also be seen in certain Painlev\'e transcendent evaluations of the
gap probabilities \cite{Fo06}. These expressions are given in terms
of the solution of the $\sigma$-form of the $P_{\rm III'}$ equation
\begin{equation}\label{1.6a}
(t \sigma '')^2 - v_1 v_2 (\sigma')^2 + \sigma'(4 \sigma' - 1)
(\sigma - t \sigma') - {1 \over 4^3} (v_1 - v_2)^2 = 0
\end{equation}
with
$$
v_1 = v_2 = a = \pm {1 \over 2}
$$
subject to the boundary condition
\begin{equation}\label{1.6b}
\sigma(t;a) \: \mathop{\sim}\limits_{t \to 0^+}  \: {t^{1 + a} \over
2^{2 + 2a} \Gamma(1 + a) \Gamma(2 + a) }.
\end{equation}
In terms of this solution, introduce the corresponding
$\tau$-functions by
\begin{equation}\label{1.6c}
\tau_{\rm III'}(s;a) := \exp \Big ( - \int_0^s {\sigma(t;a) \over t}
\, dt \Big ).
\end{equation}
Then
\begin{align}
E_1^{\rm bulk}(0;(0,2s)) & =  \tau_{\rm III'}((\pi s)^2;-1/2)
\label{t1}
\\
E_2^{\rm bulk}(0;(0,2s)) & =  \tau_{\rm III'}((\pi s)^2;-1/2)
\,\tau_{\rm III'}((\pi s)^2;1/2) \label{t2}
\\
E_4^{\rm bulk}(0;(0,2s)) & =  {1 \over 2} \Big ( \tau_{\rm
III'}((\pi s)^2;-1/2) + \tau_{\rm III'}((\pi s)^2;1/2) \Big ).
\label{t3}
\end{align}
Comparison of the results (\ref{S1})--(\ref{S3}) with the results
(\ref{t1})--(\ref{t3}) shows
\begin{equation}\label{g1}
\det  ( \mathbb{I} - K_{(0,2s)}^{\rm bulk, \pm}) = \tau_{\rm
III'}((\pi s)^2; \mp 1/2).
\end{equation}

It is the objective of this paper to give formulas analogous to
(\ref{g1}) for both the soft and hard edge scalings. In so doing we
will be relating known $\tau$-function evaluations of these
quantities to some recently derived Fredholm determinant formulas in
the case of the soft edge, and to some new Fredholm determinant
formulas in the case of the hard edge. Further, these identities
will be generalized to include a generating function type parameter
$\xi$.

\section{Soft edge scaling}
\setcounter{equation}{0} Soft edge scaling refers to shifting the
origin to the neighbourhood of the largest, or smallest, eigenvalue
where it is required that the support of the eigenvalue density is
unbounded beyond this eigenvalue, and then scaling so that the
average eigenvalue spacings in this neighbourhood are of order
unity.

The soft edge scaling can be made precise in the case of the
Gaussian and Laguerre ensembles. For this let us define a random
matrix ensemble by its eigenvalue p.d.f., assumed to be of the
functional form
\begin{equation}\label{g}
{1 \over C} \prod_{l=1}^N g(x_l) \prod_{1 \le j < k \le N} |x_k -
x_j|^\beta,
\end{equation}
and denote the corresponding probability that the interval $J$ is
free of eigenvalues by $E_\beta(0;J;g(x);N)$. For the Gaussian
ensembles with $\beta = 1$ or 2, the soft edge scaling is defined by
\begin{equation}\label{1G}
E_\beta^{\rm soft}(0;(s,\infty))  :=  \lim_{N \to \infty} E_\beta
\Big ( 0;(\sqrt{2N} + {s \over \sqrt{2} N^{1/6} }, \infty);
e^{-\beta x^2/2};N \Big )
\end{equation}
while for $\beta = 4$ a more natural definition (see the formulas of
\cite{AFNV00}) is
\begin{equation}\label{4G}
E_4^{\rm soft}(0;(s,\infty))  :=  \lim_{N \to \infty} E_4 \Big (
0;(\sqrt{2N} + {s \over \sqrt{2} N^{1/6} }, \infty); e^{- x^2};N/2
\Big )
\end{equation}
It is expected that for a large class of weights $g(x)$ in
(\ref{g}), the soft edge limit of the gap probabilities exists and
is equal to that for the Gaussian ensembles (see \cite{De99} for
some proofs related to this statement). This can be checked
explicitly in the case of the Laguerre ensembles (i.e.~the weight
$g(x) = x^a e^{-x}$, $x>0$ in (\ref{g}), up to scaling of $x$). Thus
for $\beta = 1$ or 2 we have
\begin{equation}\label{1L}
\lim_{N \to \infty} E_\beta \Big ( 0;(4N + 2 (2N)^{1/3} s, \infty);
x^a e^{-\beta x/2};N \Big ) = E_\beta^{\rm soft}(0;(s,\infty))
\end{equation}
while for $\beta = 4$
\begin{equation}\label{1L1}
\lim_{N \to \infty} E_4 \Big ( 0;(4N + 2 (2N)^{1/3} s, \infty); x^a
e^{- x};2N \Big ) = E_4^{\rm soft}(0;(s,\infty)).
\end{equation}

A number of exact expressions are known for the $E_\beta^{\rm
soft}$. Let us consider first those involving Painlev\'e
transcendents. These can in turn be grouped into two types. The
first of these relates to the particular Painlev\'e II transcendent
$q(s)$, specifed as the solution of the Painlev\'e II equation
\begin{equation}\label{2.50}
q'' = s q + 2 q^3 + \alpha
\end{equation}
with $\alpha = 0$ and subject to the boundary condition
\begin{equation}\label{2.5a}
q(s) \mathop{\sim}\limits_{s \to \infty} {\rm Ai}(s)
\end{equation}
where ${\rm Ai}(s)$ denotes the Airy function. One has
\cite{TW94a,TW96} (see \cite{Fo00} for a simplified derivation of
the latter two)
\begin{align}
E_2^{\rm soft}(0;(s,\infty)) & =  \exp \Big ( - \int_s^\infty (t - s) q^2(t) \, dt \Big ) \\
E_1^{\rm soft}(0;(s,\infty)) & =  \exp \Big ( - {1 \over 2}
\int_s^\infty (t - s) q^2(t) \, dt \Big ) \exp \Big ( {1 \over 2}
\int_s^\infty q(t) \, dt \Big )
\label{2.7}\\
E_4^{\rm soft}(0;(s,\infty)) & =  {1 \over 2} \exp \Big ( - {1 \over
2} \int_s^\infty (t - s) q^2(t) \, dt \Big )
\nonumber \\
& \qquad \times \Big ( \exp \Big ( {1 \over 2} \int_s^\infty q(t) \,
dt \Big ) +
 \exp \Big ( - {1 \over 2} \int_s^\infty q(t) \, dt \Big )  \Big ).
\end{align}

The alternative  Painlev\'e expressions relate to the $\sigma$-form
of the P${}_{\rm II}$ equation
\begin{equation}\label{2.9b}
(H_{II}'')^2 + 4 (H_{II}')^3 + 2 H_{II}'(t H_{II}' - H_{II} ) - {1
\over 4} ( \alpha + {1 \over 2} )^2 = 0.
\end{equation}
Introduce the auxiliary Hamiltonian
\begin{equation}\label{hH}
h_{II}(t;\alpha) := H_{II}(t;\alpha) + {t^2 \over 8}
\end{equation}
and the corresponding $\tau$-function
\begin{equation}\label{th}
\tau_{II}(s;\alpha) = \exp \Big ( - \int_s^\infty h_{II}(t;\alpha)
\, dt \Big ).
\end{equation}
Then from \cite{FW02} we know that
\begin{align}
E_1^{\rm soft}(0;(s,\infty)) & = \tau_{II}^+(s;0) \label{2.11a} \\
E_2^{\rm soft}(0;(s,\infty)) & =  \tau_{II}^+(s;0) \tau_{II}^-(s,0)  \label{2.11b} \\
E_4^{\rm soft}(0;(s,\infty)) & =  {1 \over 2} \Big (
\tau_{II}^+(s;0) +  \tau_{II}^-(s;0) \Big ) \label{E12}
\end{align}
where $\tau_{II}^{\pm}(s,0)$ is specified by (\ref{th}) with
$h_{II}(t;0)$ in (\ref{hH}) subject to the boundary condition
$h_{II}(t;0) \sim \pm {1 \over 2} {\rm Ai}(t)$ as $t \to \infty$.

We turn our attention now to Fredholm determinant expressions for
the gap probabilities at the soft edge. The best known is the $\beta
= 2$ result \cite{Fo93}
\begin{equation}\label{A}
E_2^{\rm soft}(0;(s,\infty)) = \det (\mathbb{I} - K^{\rm
soft}_{(s,\infty)} )
\end{equation}
where $ K^{\rm soft}_{(s,\infty)}$ is the integral operator on
$(s,\infty)$ with kernel
\begin{equation}
K^{\rm soft}(x,y) = { {\rm Ai }(x) {\rm Ai}'(y) -  {\rm Ai }(y)
{\rm Ai}'(x) \over x - y}.
\end{equation}
This can be rewritten \cite{TW94a}
\begin{equation}\label{2.14}
E_2^{\rm soft}(0;(s,\infty)) = \det (\mathbb{I} - \tilde{K}^{\rm
soft}_{(0,\infty)} )
\end{equation}
where $ \tilde{K}^{\rm soft}_{(0,\infty)}$ is the integral operator
on $(0,\infty)$ with kernel
$$
\tilde{K}^{\rm soft} = \int_0^\infty {\rm Ai}(s+x+t) {\rm Ai}(s+y+t)
\, dt,
$$
which in turn implies
\begin{equation}\label{2V}
E_2^{\rm soft}(0;(s,\infty)) = \det ( \mathbb{I} - V^{\rm
soft}_{(0,\infty)} ) \det ( \mathbb{I} + V^{\rm soft}_{(0,\infty)} )
\end{equation}
where $V^{\rm soft}_{(0,\infty)}$ is the integral operator on
$(0,\infty)$ with kernel
\begin{equation}
 V^{\rm soft}(x,u) = {\rm Ai}(x+u+s).
\end{equation}

Recently it has been conjectured by Sasamoto \cite{Sa}, and
subsequently proved by Ferrari and Spohn \cite{FS05} that
\begin{equation}\label{2V1}
E_1^{\rm soft}(0;(s,\infty)) = \det (\mathbb{I} -  V^{\rm
soft}_{(0,\infty)} ),
\end{equation}
which is the soft edge analogue of the evaluation of $E_1^{\rm
bulk}(0;(0,s))$ (\ref{S1}). Comparing (\ref{2V}), (\ref{2V1}) with
(\ref{2.11b}), we see immediately that
\begin{equation}\label{2.23}
\tau_{II}^{\pm}(s;0) = \det (\mathbb{I} \mp V^{\rm
soft}_{(0,\infty)} ).
\end{equation}
This is the soft edge analogue of the bulk identity (\ref{g1}).

\section{Hard edge scaling}
\setcounter{equation}{0} The Laguerre ensemble has its origin in
positive definite matrices $X^\dagger X$ where $X$ is an $n \times
N$ matrix $(n \ge N)$ with real $(\beta = 1)$, complex $(\beta = 2$)
or real quaternion $(\beta = 4)$ entries. Being positive definite
the eigenvalue density is strictly zero for $x<0$; for this reason
the neighbourhood of $x=0$ is referred to as the hard edge. The hard
edge scaling limit takes $N \to \infty$ while keeping the mean
spacing between eigenvalues near $x=0$ of order unity. In relation
to the gap probabilities, this can be accomplished by the limits
$$
E_\beta^{\rm hard}(0;(0,s);a) := \lim_{N \to \infty}
E_\beta\left(0;(0, {s \over 4 N}); x^a e^{-\beta x /2}; N\right)
$$
for $\beta = 1,2$, while for $\beta = 4$
$$
E_4(0;(0,s);a) := \lim_{N \to \infty} E_4 \left(0;(0, {s \over 4N});
x^a e^{- x };  N/2\right).
$$

As for the soft edge, there are two classes of Painlev\'e
evaluations of the gap probability at the hard edge. The first
involves the solution $\tilde{q}(t)$ of the nonlinear equation
\begin{equation}\label{2.63}
t (\tilde{q}^2 - 1) (t \tilde{q}')' = \tilde{q} (t \tilde{q}')^2 +
{1 \over 4} (t - a^2) \tilde{q} + {1 \over 4} t
\tilde{q}^3(\tilde{q}^2 - 2)
\end{equation}
(a transformed version of the Painlev\'e V equation) subject to the
boundary condition
\begin{equation}\label{3.1b}
\tilde{q}(t;a) \: \mathop{\sim}\limits_{t \to 0^+} \: {1 \over 2^a
\Gamma(1+a)} t^{a/2}.
\end{equation}
Thus \cite{TW94b}
\begin{equation}
E_2^{\rm hard}(0;(0,s);a) = \exp \Big ( - {1 \over 4} \int_0^s \Big
( \log {s \over t} \Big ) \tilde{q}^2(t;a) \, dt \Big )
\end{equation}
while \cite{Fo00}
\begin{align}
E_1^{\rm hard}(0;(0,s);{\textstyle {a-1 \over 2}})  &=
 \exp \Big ( - {1 \over 8} \int_0^s \Big ( \log {s \over t}
\Big ) \tilde{q}^2(t;a) \, dt \Big ) \exp \Big ( - {1 \over 4}
\int_0^s {\tilde{q}(t;a) \over \sqrt{t} } \, dt \Big )
\label{3.3}\\
E_4^{\rm hard}(0;(0,s);a+1)
 &={1 \over 2}
 \exp \Big ( - {1 \over 8} \int_0^s \Big ( \log {s \over t}
\Big ) \tilde{q}^2(t;a) \, dt \Big ) \nonumber \\
&\qquad \times \Big ( \exp \Big ( - {1 \over 4} \int_0^s
{\tilde{q}(t;a) \over \sqrt{t} } \, dt \Big ) + \exp \Big ( {1 \over
4} \int_0^s {\tilde{q}(t;a) \over \sqrt{t} } \, dt \Big ) \Big ).
\end{align}

For the second class of Painlev\'e evaluations at the hard edge, we
recall the $\sigma$-form of the $P_V$ equation
\begin{multline}\label{sV}
(t \sigma'')^2 - \Big ( \sigma - t \sigma' + 2 (\sigma')^2 +
(\nu_0 + \nu_1 + \nu_2 + \nu_3) \sigma' \Big )^2
\\ + 4(\nu_0 + \sigma') (\nu_1 + \sigma') (\nu_2 + \sigma') (\nu_3 +
\sigma') = 0.
\end{multline}
Set
\begin{equation}\label{sV1}
\nu_0 = 0, \quad \nu_1 = v_2 - v_1, \quad \nu_2 = v_3 - v_1, \quad
\nu_3 = v_4 - v_1
\end{equation}
and let
\begin{equation}\label{3.6a}
x \tilde{h}_V^{\pm}(x;a) = \sigma^{\pm}(x;a) - {1 \over 4} x^2 +
{a-1 \over 2} x - {a (a -1) \over 4}
\end{equation}
where $\sigma^{\pm}(x;a)$ satisfies (\ref{sV}) with $t \mapsto 2x$,
subject to the boundary condition consistent with
\begin{equation}\label{3.7}
x \tilde{h}_V^{\pm}(x;a)
 \: \mathop{\sim}\limits_{x \to 0^+} \: \mp {x^{a+1} \over 2^{a+1} \Gamma(a+1) }.
\end{equation}
Further, introduce the $\tau$-function
\begin{equation}
\tau_V^{\pm}(s;a) = \exp \int_0^s  \tilde{h}_V^{\pm}(x;a) \, dx.
\end{equation}
In terms of this quantity \cite{FW02}
\begin{align}
E_1^{\rm hard}(0;(0,s); {a - 1 \over 2} ) & = \tau_V^+(\sqrt{s};a) \label{3.9y} \\
E_2^{\rm hard}(0;(0,s); a ) & =  \tau_V^+(\sqrt{s};a) \tau_V^-(\sqrt{s};a) \label{3.9x} \\
E_4^{\rm hard}(0;(0,s); a +1 ) & =  {1 \over 2} \Big (
\tau_V^+(\sqrt{s};a) + \tau_V^-(\sqrt{s};a) \Big ) \label{3.9}
\end{align}
where the parameters (\ref{sV1}) are specified by
$$
v_1 = - v_3 = - (a-1)/4, \qquad v_2 = - v_4 = (a+1)/4.
$$

In relation to Fredholm determinant expressions for the gap
probabilities at the soft edge, analogous to (\ref{A}) we have
\cite{Fo93}
\begin{equation}\label{3.xx}
E_2^{\rm hard}((0,s);a) = \det (\mathbb{I} - K^{\rm hard}_{(0,s)})
\end{equation}
where $ K^{\rm hard}_{(0,s)}$ is the integral operator on $(0,s)$
with kernel
\begin{equation}\label{hardkernel}
K^{\rm hard}(x,y) = {J_a(\sqrt{x}) \sqrt{y} J_a'(\sqrt{y}) -
\sqrt{x} J_a'(\sqrt{x}) J_a(\sqrt{y}) \over x - y}.
\end{equation}
This can be rewritten \cite{TW94b}
\begin{equation}\label{3.12}
E_2^{\rm hard}((0,s);a) = \det (\mathbb{I} - \tilde{K}^{\rm
hard}_{(0,1)})
\end{equation}
where $ \tilde{K}^{\rm hard}_{(0,1)}$ is the integral operator on
$(0,1)$ with kernel
\begin{equation}
 \tilde{K}^{\rm hard}(x,y) = {s \over 4} \int_0^1 J_a(\sqrt{sxu}) J_a(\sqrt{syu}) \, du.
\end{equation}
Because
\begin{equation}\label{3.19}
 \tilde{K}^{\rm hard}_{(0,1)} = (V_{(0,1)}^{\rm hard})^2
\end{equation}
where $V_{(0,1)}^{\rm hard}$ is the integral operator on $(0,1)$
with kernel
\begin{equation}\label{defV}
V^{\rm hard}(x,y) = {\sqrt{s} \over 2} J_a(\sqrt{sxy}),
\end{equation}
it follows that
\begin{equation}\label{3.20}
E_2^{\rm hard}((0,s);a) = \det (\mathbb{I} - V_{(0,1)}^{\rm hard})
\det  (\mathbb{I} + V_{(0,1)}^{\rm hard}).
\end{equation}

For $\beta = 1$, a Fredholm determinant expression analogous to the
result (\ref{2V1}) holds true.  This is proved with the help of the
three following lemmas, which are modeled on the strategy used in
\cite{FS05} to prove (\ref{2V1}).

\begin{lemma}\label{L1} Let $V=V_{(0,1)}^{\rm hard}$ and $\rho(x)=1/\sqrt{x}$ for $x>0$.  Let $\langle
f|g\rangle_{(0,1)}=\int_0^1f(x)g(x)dx$ be the scalar product in
$\mathfrak{L}^2(0,1)$.  Let also $\delta_1$ denote the delta
function at $1$; that is, $\langle \delta_1|f\rangle_{(0,1)}=f(1)$.
Then,
$$
\left(E_1^{\rm
hard}\left((0,s);{\textstyle\frac{a-1}{2}}\right)\right)^2=\det
(\mathbb{I} - V)\det (\mathbb{I} + V)\langle\delta_1|(\mathbb{I}
+V)^{-1}\rho\rangle_{(0,1)}.
$$
\end{lemma}
\begin{proof}
We know from \cite{Fo00} that
$$\left(E_1^{\rm
hard}\left((0,s);{\textstyle\frac{a-1}{2}}\right)\right)^2=\det
\left(\mathbb{I}-K^{\mathrm{hard}}_{(0,s)} -C\otimes D\right),$$
where $K^{\mathrm{hard}}_{(0,s)}$ and $C\otimes D$  are integral
operators on $(0,s)$ whose kernels are respectively
$K^{{\mathrm{hard}}}(x,y)$ (see Eq.~(\ref{hardkernel})) and
$J_a(\sqrt{x})\frac{1}{2\sqrt{y}}\int_{\sqrt{y}}^\infty J_a(t)dt$.
Note that $f\otimes g$ stands for an integral operator with kernel
\begin{equation}\label{tensprod}(f\otimes
g)(x,y)=f(x)g(y).\end{equation}We now make use of
$\sqrt{s}J_a(\sqrt{x})=2(V\delta_1)(x)$ and $\int_0^\infty
J_a(y)dy=1$ for showing that
\begin{align}
(C\otimes Df)(x)&=
\frac{J_a(\sqrt{x})}{2}\int_0^s\left(1-\int_0^{\sqrt{y}}J_a(t)dt\right)\frac{f(y)}{\sqrt{y}}dy\nonumber\\
&=\frac{\sqrt{s}}{2}J_a(\sqrt{x})
\int_0^1\left(\frac{1}{\sqrt{y}}-\frac{\sqrt{s}}{2}\int_0^1\frac{J_a(\sqrt{syt})}{\sqrt{t}}dt
\right)f(sy)dy\nonumber\\
&=(V\delta_1)(x)\int_0^1\Big(\rho(y)-(V\rho)(y)\Big)f(sy)dy.
\nonumber
\end{align}
Then by recalling Eqs \eqref{3.xx}--\eqref{3.19}, we  get
\begin{align} \left(E_1^{\rm
hard}\left((0,s);{\textstyle\frac{a-1}{2}}\right)\right)^2&=\det
\left(\mathbb{I}-V^2
-V\delta_1\otimes(\mathbb{I} -V)\rho\right)\nonumber\\
&=\det(\mathbb{I} -V)\det(\mathbb{I} +V)\det\left(\mathbb{I}
-(\mathbb{I} +V)^{-1}\rho\otimes V\delta_1\right)\label{blabla}
\end{align}
 But $\mathbb{I} -(\mathbb{I} +V)^{-1}\rho\otimes V\delta_1$ is a degenerate operator of
rank $1$ (see e.g.\ \cite[Eq.\ (17)]{TW96}). This means that
Eq.~\eqref{blabla} can be written as
$$ \left(E_1^{\rm
hard}\left((0,s);{\textstyle\frac{a-1}{2}}\right)\right)^2=\det(\mathbb{I}
-V)\det(\mathbb{I} +V)\left(1 -\langle\delta_1|(\mathbb{I}
+V)^{-1}V\rho\rangle_{(0,1)}\right).$$The use of
$\langle\rho|\delta_1\rangle=1$ finishes the proof.
\end{proof}

\begin{lemma}\label{L2}
Let $\Delta$ be the operator defined by $(\Delta f)(x)=x\partial_x
f(x)$ and let $\otimes$ be the direct product defined in
Eq.~(\ref{tensprod}).  Then, for $V=\BB$,
$$ 2s\frac{\partial}{\partial s}V=(\mathbb{I} +2\Delta)V,\qquad
\Delta V=-V\Delta+V\delta_1\otimes\delta_1-V,$$  and consequently,
$$s\frac{\partial}{\partial s}V=\frac{1}{2}(\mathbb{I} -V^2)^{-1}V(\mathbb{I} +2\Delta)-(\mathbb{I} -V^2)^{-1}V\delta_1\otimes\delta_1(\mathbb{I} +V)^{-1}.$$
\end{lemma}
\begin{proof}
Firstly, the definition of $V=\BB$ ( given in Eq.~\eqref{defV}) and
the property
$s\partial_sJ_{a}(\sqrt{sxt})=x\partial_xJ_{a}(\sqrt{sxt})$ directly
imply that$$ s\frac{\partial}{\partial s}(Vf)(x)=
s\frac{\partial}{\partial
s}\left(\frac{\sqrt{s}}{2}\int_0^1J_{a}(\sqrt{sxt})f(t)dt\right)=\frac{1}{2}(Vf)(x)+(\Delta
Vf)(x)$$ which is the desired result.  Secondly, by using
$x\partial_xJ_{a}(\sqrt{sxt})=t\partial_tJ_{a}(\sqrt{sxt})$ and by
integrating by parts, we find
\begin{align} (\Delta
Vf)(x)&=\frac{\sqrt{s}}{2}\int_0^1t\frac{\partial}{\partial
t}\left(J_{a}(\sqrt{sxt})\right)f(t)dt\nonumber\\
&=\frac{\sqrt{s}}{2}J_a(\sqrt{sx})f(1)-\frac{\sqrt{s}}{2}\int_0^1J_{a}(\sqrt{sxt})f(t)dt\nonumber\\
&\qquad\qquad-\frac{\sqrt{s}}{2}\int_0^1J_{a}(\sqrt{sxt})J_{a}(\sqrt{sxt})t\frac{\partial}{\partial
t}\left(f(t)\right)dt\nonumber\\
&=(V\delta_1)(x)\langle\delta_1|f\rangle_{(0,1)}-(Vf)(x)-(V\Delta
f)(x),\nonumber\end{align} as expected.  Finally, by exploiting
$2s{\partial_s}V=(\mathbb{I}+2\Delta)V$, $(\mathbb{I} +V)^{-1}=
\sum_{n\geq0}(-1)^n V^n$ and $(\mathbb{I}
+V)^{-2}=\sum_{n\geq0}(-1)^n(n+1) V^n$, we get
\begin{align}
2s\frac{\partial}{\partial s}(\mathbb{I} +V)^{-1}&:=\sum_{n\geq
1}(-1)^n
s\frac{\partial}{\partial s}V^n\nonumber\\
&=\sum_{n\geq 1}(-1)^n \sum_{k=0}^{n-1}V^k
\left(2s\frac{\partial}{\partial
s}V\right)V^{n-k-1}\nonumber\\
&=-V(\mathbb{I} +V)^{-2}+2\sum_{n\geq 1}(-1)^n \sum_{k=0}^{n-1}V^k
\Delta V^{n-k},\nonumber
\end{align}
 But,
for any operators $O$ and $P$ such that $OV=-VO-P$, we have
\cite[Lemma 3]{FS05}
$$
\sum_{n\geq 1}(-1)^n \sum_{k=0}^{n-1}V^k O V^{n-k}=(\mathbb{I}
-V^2)^{-1}VO+(\mathbb{I} -V^2)^{-1}P(\mathbb{I} +V)^{-1}.
$$In our case, $O=\Delta$ and $P=-V\delta_1\otimes\delta_1+V$.
Therefore,
\begin{multline*}2s\frac{\partial}{\partial s}(\mathbb{I}
+V)^{-1}=-V(\mathbb{I} +V)^{-2}+2(\mathbb{I} -V^2)^{-1}V(\mathbb{I}
+V)^{-1}\\+2(\mathbb{I} -V^2)^{-1}V\Delta-(\mathbb{I}
-V^2)^{-1}V\delta_1\otimes\delta_1(\mathbb{I}
+V)^{-1}.\end{multline*} This turns out to be equivalent to the last
equation we wanted to prove.
\end{proof}

\begin{lemma}\label{L3}  Let  $M$ be a symmetric,
trace class operator in $\mathfrak{L}^2(0,1)$.  Then,
$$\mathrm{Tr}\left[(\mathbb{I} +2\Delta)M\right]=\langle\delta_1|M\delta_1\rangle_{(0,1)}.$$
\end{lemma}
\begin{proof} Set $\{f_i\}$ and $\{\lambda_i\}$, respectively the
orthonormal eigenfunctions and the eigenvalues of $M$. On the one
hand, we have
$$\langle\delta_1|M\delta_1\rangle_{(0,1)}=\sum_i\lambda_if_i(1)^2.$$
 On the other hand, we have
$$\mathrm{Tr}[(\mathbb{I} +2\Delta)M]=\sum_i\langle
f_i|(1+2\Delta)Mf_i\rangle_{(0,1)}=\sum_i\lambda_i\left(1+2\langle
f_i|\Delta f_i\rangle_{(0,1)}\right).$$ But integration by parts
gives
$$\langle
f_i|\Delta
f_i\rangle_{(0,1)}=\int_0^1f_i(x)x\frac{\partial}{\partial
x}f_i(x)dx=f_i(1)^2-1-\int_0^1f_i(x)x\frac{\partial}{\partial
x}f_i(x)dx.$$ Consequently, $ \mathrm{Tr}[(\mathbb{I}
+2\Delta)M]=\sum_i\lambda_i f_i(1)^2$ and the lemma follows.
\end{proof}

\begin{prop}
We have
\begin{equation}\label{3.18}
E_1^{\rm hard}\left((0,s);{a-1 \over 2}\right) = \det (\mathbb{I}  -
V_{(0,1)}^{\rm hard}),
\end{equation}and consequently $$ \tau_V^+(\sqrt{s})=\det (\mathbb{I}  -
V_{(0,1)}^{\rm hard}).$$
\end{prop}
\begin{proof} From Lemma \ref{L1}, we know that the proposition is true if
\begin{equation}
\label{3.22a} \det\left((\mathbb{I} -V)(\mathbb{I} +V)^{-1}\right)=
\langle\rho|(\mathbb{I} +V)^{-1}\delta_1\rangle_{(0,1)}
\end{equation}
or equivalently, if
\begin{equation} \label{PE1}\ln\det\left((\mathbb{I} -V)(\mathbb{I} +V)^{-1}\right)= \ln
\langle\delta_1|(\mathbb{I}
+V)^{-1}\rho\rangle_{(0,1)}.\end{equation}But from the fact that
$V\rightarrow0$ as $s\rightarrow0$, we deduce that Eq.~\eqref{PE1}
holds if and only if
$$
 s\frac{\partial}{\partial s}\ln\det\left((\mathbb{I} -V)(\mathbb{I} +V)^{-1}\right)= s\frac{\partial}{\partial s}\ln
\langle\delta_1|(\mathbb{I} +V)^{-1}\rho\rangle_{(0,1)}.$$By virtue
of $s\partial_s \ln (\det M)=\mathrm{Tr}(M^{-1}s\partial_sM)$, the
latter equation reads
\begin{equation}\label{PE2}
\mathrm{Tr}\left[(\mathbb{I} -V^2)^{-1}2 s\frac{\partial}{\partial
s}V\right]=
 -\frac{\langle\delta_1|s\frac{\partial}{\partial s}(\mathbb{I} +V)^{-1}\rho\rangle_{(0,1)}}{
\langle\delta_1|(\mathbb{I}
+V)^{-1}\rho\rangle_{(0,1)}}.\end{equation}Using the cyclicity of
the trace and Lemma \ref{L3}, we find that
\begin{equation}\label{PE3}
\mathrm{Tr}\left[(\mathbb{I} -V^2)^{-1}2 s\frac{\partial}{\partial
s}V\right]=\mathrm{Tr}\left[(\mathbb{I} -V^2)^{-1}(\mathbb{I}
+2\Delta)V\right]=\langle\delta_1|(\mathbb{I}
-V^2)^{-1}V\delta_1\rangle_{(0,1)}.\end{equation} Furthermore, Lemma
\ref{L2} and $(\mathbb{I} +2\Delta)\rho=0$ imply that
\begin{multline}\label{PE4}
-\frac{\langle\delta_1|s\frac{\partial}{\partial s}(\mathbb{I}
+V)^{-1}\rho\rangle_{(0,1)}}{ \langle\delta_1|(\mathbb{I}
+V)^{-1}\rho\rangle_{(0,1)}}\\=\frac{\langle\delta_1|(\mathbb{I}
-V^2)^{-1}V\delta_1\otimes\delta_1(\mathbb{I}
+V)^{-1}\rho\rangle_{(0,1)}}{ \langle\delta_1|(\mathbb{I}
+V)^{-1}\rho\rangle_{(0,1)}}=\langle\delta_1|(\mathbb{I}
-V^2)^{-1}V\delta_1\rangle_{(0,1)}.\end{multline} The comparison of
Eqs (\ref{PE3})--(\ref{PE4}) finally establishes the validity of
Eq.~(\ref{PE2}), and  the proposition follows.\end{proof}

By comparing (\ref{3.18}) with (\ref{3.9y}), and then equating
(\ref{3.9x}) and (\ref{3.20}), we obtain the hard edge analogue of
(\ref{2.23}).

\begin{cor}
One has
\begin{equation}\label{3.23}
\tau_V^{\pm}(\sqrt{s}) = \det (\mathbb{I} \mp V_{(0,1)}^{\rm hard}).
\end{equation}
\end{cor}

We remark that the evaluation of the hard edge gap probability
(\ref{3.18}), and the identity  (\ref{3.23}), contain the
evaluation of the soft edge gap probability (\ref{2V1}), and the
identity (\ref{2.23}), as a limiting case. This follows from the
limit formula (see e.g.~\cite{BF03}),
$$
E_1^{\rm soft}(0;(s,\infty)) = \lim_{a \to \infty} E_1^{\rm
hard}\left(0;(0,a^2-(2a^2)^{2/3}s); {a - 1 \over 2} \right).
$$

\section{Generating function generalization}
\setcounter{equation}{0} The probabilistic quantity $E_2^{\rm
bulk}(0;(0,s))$ is the first member of the sequence $\{E_2^{\rm
bulk}(n;(0,s))\}_{n=0,1,\dots}$ where $E_2^{\rm bulk}(n;(0,s))$
denotes the probability that the interval $(0,s)$ contains exactly
$n$ eigenvalues. Introducing the generating function for this
sequence by
\begin{equation}\label{4.0}
E_2^{\rm bulk}((0,s);\xi) := \sum_{n=0}^\infty (1 - \xi)^n E_2^{\rm
bulk}(n;(0,s)),
\end{equation}
it is well known that \cite{Gaudin}
\begin{eqnarray}\label{4.1}
E_2^{\rm bulk}((0,s);\xi) & = & \det (\mathbb{I} - \xi K_{(0,s)}^{\rm bulk}) \nonumber \\
& = & \det (\mathbb{I} - \xi K_{(0,s)}^{\rm bulk,+})\det (\mathbb{I}
- \xi K_{(0,s)}^{\rm bulk,-}).
\end{eqnarray}
Thus to obtain from the Fredholm determinant expressions (\ref{1.5})
for $E_2^{\rm bulk}(0;(0,s))$ expressions for the generating
function (\ref{4.0}), one merely multiplies the kernel by $\xi$.

This immediately raises the question as to whether the formula
(\ref{g1}) admits a generalization upon multiplying the kernel by
$\xi$? The answer is that it does, with the only change being in the
initial condition (\ref{1.6b}) satisfied by the transcendent
$\sigma(t;a)$ in (\ref{1.6c}). Thus specify $\sigma(t;a)$ as again
satisfying (\ref{1.6a}), but now subject to the boundary condition
$$
\sigma(t;a;\xi) \: \mathop{\sim}\limits_{t \to 0^+}  \: {\xi t^{1 +
a} \over 2^{2 + 2a} \Gamma(1 + a) \Gamma(2 + a) }.
$$
Then with
$$
\tau_{\rm III'}(s;a;\xi) := \exp \Big ( - \int_0^s {\sigma(t;a;\xi)
\over t} \, dt \Big )
$$
we have \cite{TW94b,Fo06}
\begin{equation}\label{4.2a}
 \det (\mathbb{I}- \xi K_{(0,2s)}^{\rm bulk,\pm}) = \tau_{\rm III'}((\pi s)^2, \mp 1/2; \xi).
\end{equation}

Now, the gap probabilities at the soft and hard edges can similarly
be generalized to generating functions. Thus, in an obvious notation
\begin{align*}
E_2^{\rm soft}((s,\infty);\xi) & = \sum_{n=0}^\infty (1 - \xi)^n
E_2^{\rm soft}(n;(s,\infty)) \\
E_2^{\rm hard}((0,s);a;\xi) & =  \sum_{n=0}^\infty (1 - \xi)^n
E_2^{\rm hard}(n;(0,s);a).
\end{align*}
Analogous to (\ref{4.1}), it is fundamental in random matrix theory
that (\ref{2.14}) and (\ref{3.12}) generalize (see e.g.~\cite{Fo02})
to give
\begin{align}\label{4.4}
E_2^{\rm soft}((s,\infty);\xi) & =  \det (\mathbb{I}- \xi
\tilde{K}^{\rm soft}_{(0,\infty)} )
\nonumber \\
& =  \det ( \mathbb{I} - \sqrt{\xi} V^{\rm soft}_{(0,\infty)} ) \det
( \mathbb{I} + \sqrt{\xi} V^{\rm soft}_{(0,\infty)} )
\end{align}
and
\begin{align}\label{4.5}
E_2^{\rm hard}((0,s);a) & =  \det (\mathbb{I} - \xi \tilde{K}^{\rm
hard}_{(0,1)})
\nonumber \\
& =
 \det (\mathbb{I} - \sqrt{\xi} V_{(0,1)}^{\rm hard}) \det  (\mathbb{I} + \sqrt{\xi} V_{(0,1)}^{\rm hard}).
\end{align}

Also, analogous to the situation with $E_2^{\rm bulk}((0,s);\xi)$ we
know from \cite{TW94a,TW94b,FW02} that the $\tau$-function formulas
in (\ref{E12}) and (\ref{3.9x}) for $E_2^{\rm soft}(0;(s,\infty))$
and $E_2^{\rm hard}(0;(0,s))$ require only modification to the
boundary condition satisfied by the corresponding transcendent to
generalize to $\tau$-function formulas for the generating functions.
Explicitly, in relation to $E_2^{\rm soft}$, in (\ref{2.9b}) and
(\ref{hH}) again set $\alpha = 0$, but now require that $H_{\rm II}$
and thus $h_{\rm II}$ depend on an auxiliary parameter $\xi$ by
specifying the boundary condition
\begin{equation}\label{Ap}
h_{\rm II}^\pm(t;0;\xi) \mathop{\sim}\limits_{t \to \infty} \pm
{\sqrt{\xi} \over 2} {\rm Ai}(t).
\end{equation}
Then, with
$$
\tau_{\rm II}^\pm(s;\alpha;\xi) = \exp \Big ( - \int_0^\infty h_{\rm
II}^\pm(t;\alpha;\xi) \, dt \Big ),
$$
we have \cite{TW94a}
\begin{equation}\label{4.6a}
E_2^{\rm soft}((s,\infty);\xi) = \tau_{\rm II}^+(s;0;\xi) \tau_{\rm
II}^-(s;0;\xi)
\end{equation}
where the superscripts refer to the corresponding sign in
(\ref{Ap}). And generalizing the identity implied by the equality
between (\ref{2.7}) and (\ref{2.11a}) $\tau_{\rm II}^+$ admits the
further Painlev\'e transcendent form \cite{TW96, FW02}
\begin{equation}\label{4.6}
\tau_{\rm II}^{\pm}(s;0;\xi) = \exp \Big ( - {1 \over 2}
\int_s^\infty (t - s) q^2(t;\xi) \, dt \Big ) \exp \Big ( \mp {1
\over 2} \int_s^\infty q(t;\xi) \, dt \Big )
\end{equation}
where $q(t;\xi)$ satisfies (\ref{2.50}) with $\alpha = 0$ subject to
the boundary condition
\begin{equation}
q(s;\xi) \mathop{\sim}\limits_{s \to \infty} \sqrt{\xi} {\rm Ai}(s).
\end{equation}

At the hard edge again specify $\tilde{h}_V^\pm$ in terms of
$\sigma^\pm$ by (\ref{3.6a}), but now modify the boundary condition
(\ref{3.7}) by multiplying it by $\sqrt{\xi}$ and thus requiring
that
$$
x \tilde{h}_V^{\pm}(x;a;\xi)
 \: \mathop{\sim}\limits_{x \to 0^+} \: \mp {\sqrt{\xi} x^{a+1} \over 2^{a+1} \Gamma(a+1) }.
$$
With the corresponding $\tau$ function specified by
$$
\tau_V^{\pm}(s;a;\xi) = \exp \int_0^s  \tilde{h}_V^{\pm}(x;a;\xi) \,
dx
$$
we then have \cite{FW02}
\begin{equation}\label{4.8a}
E_2^{\rm hard}((0,s);a;\xi) = \tau_V^+(s;a;\xi) \tau_V^-(s;a;\xi).
\end{equation}
Analogous to (\ref{4.6}) $\tau_V^\pm$ admits the further Painlev\'e
transcendent form \cite{Fo00,FW02}
\begin{equation}\label{4.8}
\tau_V^\pm(s;a;\xi) =
 \exp \left( - {1 \over 8} \int_0^s \Big ( \log {s \over t}
\Big ) \tilde{q}^2(t;a;\xi) \, dt \right) \exp \left( \mp {1 \over
4} \int_0^s {\tilde{q}(t;a;\xi) \over \sqrt{t} } \, dt \right)
\end{equation}
where $\tilde{q}(t;a;\xi)$ satisfies (\ref{2.63}) but now with the
boundary condition
$$
\tilde{q}(t;a;\xi) \mathop{\sim}\limits_{t \to 0^+} {\sqrt{\xi}
\over 2^a \Gamma(1+a) } t^{a/2}.
$$
This with $\xi = 1$ reduces (in the "+" case) to the equality
implied by (\ref{3.9y}) and (\ref{3.3}).

The general $\xi$ bulk identity (\ref{4.2a}) leads us to investigate
if, as is true at $\xi = 1$ according to (\ref{2.23}) and
(\ref{3.23}), that the factors in the Fredholm determinant
factorizations (\ref{4.4}), (\ref{4.5}) coincide with those in the
$\tau$-function factorizations (\ref{4.6a}), (\ref{4.8a}). The
answer is that they do coincide, but to show this requires some
intermediate working. We will detail this working for the soft edge,
and be content with a sketch in the hard edge, as the strategy is
very similar.

\begin{lemma}\label{U1}
With $q(t;\xi)$ as in  (\ref{4.6})
\begin{equation}\label{La1}
\exp \Big ( - \int_s^\infty q(t;\xi) \, dt \Big ) = 1 -
\int_s^\infty [(\mathbb{I}- \xi K^{\rm soft})^{-1} A^{\rm s}](y)
B^{\rm s}(y) \, dy
\end{equation}
where $A^{\rm s}$ is the operator which multiplies by $\sqrt{\xi}
{\rm Ai}(x)$, while
\begin{equation}\label{w0}
 B^{\rm s}(y) := 1 - \sqrt{\xi} \int_y^\infty {\rm Ai}(x) \, dx.
\end{equation}
\end{lemma}
\begin{proof} We closely follow the working in \cite{Fo00}, referring to
equations therein as required. Introduce the notation
$$
\phi(x) = \sqrt{\xi} {\rm Ai}(x), \qquad Q(x) = [(\mathbb{I} - \xi
K^{\rm soft})^{-1} \phi](x)
$$
so that
\begin{equation}\label{w1}
\int_s^\infty[(\mathbb{I} - \xi K^{\rm soft})^{-1} A^{\rm s}](y)
B^{\rm s}(y) \, dy = \int_s^\infty dy \, Q(y) \Big ( 1 -
\int_y^\infty \phi(v) \, dv \Big ) =: u_\epsilon.
\end{equation}
The strategy is to derive coupled differential equations for
$u_\epsilon$ and
\begin{equation}\label{w2}
q_\epsilon := \int_s^\infty dy \, \rho(s,y)  \Big ( 1 -
\int_y^\infty \phi(v) \, dv \Big ),
\end{equation}
where $ \rho(s,y)$ denotes the kernel of the integral operator $
(\mathbb{I} - \xi K^{\rm soft})^{-1}$.

According to the working of \cite[eqs.~(3.11)--(3.14)]{Fo00} the
sought equations are
\begin{align}
{d u_\epsilon \over ds} & =  - q(s;\xi) q_\epsilon \label{x1} \\
{d q_\epsilon \over ds} & =   q(s;\xi) (1 - u_\epsilon ), \label{x2}
\end{align}
where $q(s;\xi)$ enters via the fact that $Q(s) = q(s;\xi)$. Since
$Q(y)$ is smooth while $\rho(s,y)$ is equal to the delta function
$\delta(s-y)$ plus a smooth term, we see from (\ref{w1}), (\ref{w2})
that the equations (\ref{x1}), (\ref{x2}) must be solved subject to
the boundary conditions
$$
 u_\epsilon \to 0, \qquad q_\epsilon \to 1 \qquad {\rm as} \quad s \to \infty
$$
It is simple to verify that the solution subject to these boundary
conditions is
$$
u_\epsilon(s) = 1 - q_\epsilon(s) = 1 - \exp\Big ( - \int_s^\infty
q(x;\xi) \, dx\Big ),
$$
and (\ref{La1}) follows.\end{proof}

\begin{lemma}\label{U2}
One has
\begin{equation}\label{x4}
1 - \int_s^\infty [(\mathbb{I} - \xi K^{\rm soft})^{-1} A^{\rm
s}](y) B^{\rm s}(y) \, dy = \langle \delta_0 | (\mathbb{I} +
\sqrt{\xi} V^{\rm soft}_{(0,\infty)})^{-1} 1 \rangle_{(0,\infty)}.
\end{equation}
\end{lemma}
\begin{proof}
Changing variables $y \mapsto y + s$ and noting from
(\ref{w0}) that
$$
 B^{\rm s}(y+s) = [(\mathbb{I}- \sqrt{\xi} V_{(0,\infty)}^{\rm soft})(1)](y)
$$
shows that the left hand side of (\ref{x4}) is equal to
$$
1 - \langle \delta_0 | \sqrt{\xi} V_{(0,\infty)}^{\rm soft}
(\mathbb{I} + \sqrt{\xi} V_{(0,\infty)}^{\rm soft} )^{-1} 1
\rangle_{(0,\infty)}.
$$
This reduces to the right hand side upon noting $ \langle \delta_0
|1  \rangle_{(0,\infty)}= 1$. \end{proof}

The sought $\xi$ generalization of (\ref{2.23}) can now be
established.

\begin{prop}
One has
\begin{equation}\label{T1}
\tau_{II}^{\pm}(s;0;\xi) = \det (\mathbb{I} \mp \sqrt{\xi}
V_{(0,\infty)}^{\rm soft}).
\end{equation}
\end{prop}
\begin{proof}
The well known fact \cite{TW94b} that
\begin{equation}\label{qK}
\exp \Big ( - \int_s^\infty (t - s) q^2(t;\xi) \, dt \Big ) = \det
(\mathbb{I} - \xi \tilde{K}^{\rm soft}_{(0,\infty)})
\end{equation}
together with (\ref{4.6}), Lemma \ref{U1} and Lemma \ref{U2} tell us
that
$$
(\tau_{\rm II}^+(s;0;\xi) )^2 = \det ( \mathbb{I} - \xi
\tilde{K}^{\rm soft}) \langle \delta_0 | (\mathbb{I} + \sqrt{\xi}
V_{(0,\infty)}^{\rm soft})^{-1} 1 \rangle.
$$
Recalling (\ref{4.4}) we see that (\ref{T1}) in the "+" case is
equivalent to the identity
\begin{equation}\label{V11}
\det ( \mathbb{I} -  \sqrt{\xi} V_{(0,\infty)}^{\rm soft}) =
\det(\mathbb{I} + \sqrt{\xi}   V_{(0,\infty)}^{\rm soft} )\langle
\delta_0 | (\mathbb{I} + \sqrt{\xi} V_{(0,\infty)}^{\rm soft})^{-1}
1 \rangle.
\end{equation}
With $\xi = 1$ this is precisely the identity established in
\cite{FS05}. Inspection of the details of the derivation (on which,
as already mentioned, our Lemmas \ref{L1}--\ref{L3} are based) show
that  the workings remain valid upon multiplying
$V_{(0,\infty)}^{\rm soft}$ by a scalar, so (\ref{V11}) is true, and
thus so is (\ref{T1}) in the "+" case. The validity of the "$-$"
case now follows from use of (\ref{qK}) and the plus case in
(\ref{4.4}). \end{proof}

At the hard edge, analogous to the result (\ref{T1}) we would like
to show that (\ref{3.23}) admits a $\xi$-generalization. The
$\xi$-generalization of the $\tau$-function on the left hand side is
given by (\ref{4.8a}). In relation to that expression we know that
\cite{TW94b}
$$
 \exp \Big ( - {1 \over 4} \int_0^s \Big ( \log {s \over t}
\Big ) \tilde{q}^2(t;a;\xi) \, dt \Big ) = \det (\mathbb{I} - \xi
\tilde{K}^{\rm hard}_{(0,1)})
$$
while the workings of \cite{Fo00} allow us to deduce that
\begin{equation}\label{ss1}
\exp \Big ( - {1 \over 2} \int_0^s {\tilde{q}(t;a;\xi) \over
\sqrt{t} } \, dt \Big )  = 1 - \int_0^s [(\mathbb{I} - \xi K^{\rm
hard})^{-1} A^{\rm h}](y) B^{\rm h}(y) \, dy
\end{equation}
where $A^{\rm h}$ is the operator which multiplies by
$\sqrt{\xi}J_a(\sqrt{x})$, while
$$
 B^{\rm h}(y) = {1 \over 2 \sqrt{y}} \Big ( 1 - \sqrt{\xi} \int_0^{\sqrt{y}}
J_a(t) \, dt \Big )
$$
(cf.~(\ref{La1})). Proceeding as in the proof of Lemma \ref{L1} (and
using the notation therein) shows that the right hand side of
(\ref{ss1}) is equal to
$$
\langle \delta_1 | (\mathbb{I}+ \sqrt{\xi} V_{(0,1)}^{\rm
hard})^{-1} \rho \rangle_{(0,1)}.
$$
With these preliminaries noted, our sort result can be established.

\begin{prop}
One has
\begin{equation}
\tau_V^{\pm}(s;a;\xi) = \det (\mathbb{I} \mp \sqrt{\xi}
V_{(0,1)}^{\rm hard}).
\end{equation}
\end{prop}
\begin{proof}
According to the above results, the "+" case is equivalent to
the identity
\begin{equation}\label{vv}
\det( \mathbb{I} - \sqrt{\xi} V_{(0,1)}^{\rm hard} ) = \det(
\mathbb{I} + \sqrt{\xi} V_{(0,1)}^{\rm hard} ) \langle \delta_1 |
(\mathbb{I} + \sqrt{\xi} V_{(0,1)}^{\rm hard})^{-1} \rho
\rangle_{(0,1)},
\end{equation}
which in the case $\xi=1$ is precisely (\ref{3.22a}). The derivation
given of the latter identity carries over unchanged with $V \mapsto
\sqrt{\xi} V$, thus verifying (\ref{vv}). The "minus" case can now
be deduced from (\ref{4.5}). \end{proof}

We conclude by noting a $\xi$-generalization which holds in the bulk
but not at the hard or soft edge. Thus in the bulk, with the
generating function for $\{E_1^{\rm bulk}(n;(0,s))\}_{n=0,1,\dots}$
specified by
$$
E_1^{{\rm bulk, \mp}}((0,s);\xi) = \sum_{n=0}^\infty (1 - \xi)^n
\Big ( E_1^{\rm bulk}(2n;(0,s)) + E_1^{\rm bulk}(2n \mp 1;(0,s))\Big
) ,
$$
the identity (\ref{S1}) admits the simple generalization (see
e.g.~[9])
\begin{equation}\label{f.1}
E_1^{{\rm bulk, \mp}}((0,s);\xi) = \det (\mathbb{I} + \sqrt{\xi}
K_{(0,\infty)}^{\rm bulk, \pm}).
\end{equation}
However the corresponding $\xi$ generalizations of (\ref{2V1}) and
(\ref{3.18}) cannot hold true, as the corresponding integral
operators are not positive definite, but rather have both positive
and negative eigenvalues. The Fredholm determinant $\det(\mathbb{I}
- \xi V^{\rm soft}_{(0,\infty)})$ (for example) thus vanishes for
some negative $\xi$, in contradiction to the behaviour of
$\sum_{n=0}^\infty (1 - \xi)^n E_1^{\rm soft}(n,(s,\infty))$.

\section*{Acknowledgement}
The work of P.J.F.~has been supported by the Australian Research
Council.
 P.D.~is grateful to the Natural Sciences and Engineering Research
Council of Canada for a postdoctoral  fellowship. We thank
N.S.~Witte for comments relating to \cite{FS05}.

\end{document}